\documentclass{webofc}
\usepackage[varg]{txfonts}
\usepackage{lineno}
\begin{document}

\title{CME search at STAR}

\author{\firstname{Yu} \lastname{Hu}\inst{1,2}\fnsep\thanks{\email{huy17@fudan.edu.cn}} \\
for the STAR Collaboration
    %\and
        %\firstname{Second author} \lastname{Second author}\inst{2}\fnsep\thanks{\email{Mail address for second
        %     author if necessary}} \and
        %\firstname{Third author} \lastname{Third author}\inst{3}\fnsep\thanks{\email{Mail address for last
        %     author if necessary}}
        % etc.
}

\institute{Fudan University 
\and
           Brookhaven National Laboratory
            }

\abstract{%
The hot and dense medium produced in relativistic heavy-ion collisions has been conjectured to be accompanied by an axial charge asymmetry that may lead to a separation of electric charges in the direction of the extremely strong magnetic field, also known as the Chiral Magnetic Effect (CME) \cite{Kharzeev:1999,Kharzeev:2008,Mclerran:2013}. The measurement of azimuthal correlator ($\Delta\gamma$) with respect to the spectator  plane \cite{Haojie:2018}, gives us an opportunity to measure the possible CME fraction beyond the flow background. Preliminary results using this approach with combined Au+Au collisions at $\sqrt{s_{NN}} =$ 200 GeV and U+U at $\sqrt{s_{NN}} =$ 193 GeV show $f_{CME}$ at (8±4±8)$\%$.
Meanwhile, the observability of CME has been conjectured to be dependent on $\sqrt{s_{NN}}$ due to changes in the lifetime of the magnetic field, the strengths of CME signal and non-CME background. 
At lower energies, the Event Plane Detector (EPD) installed in the year 2018 provides a unique capability for CME search. The background scenario test at Au+Au $\sqrt{s_{NN}} =$ 27 GeV by using $\Delta\gamma$ with respect to  TPC and the new installed EPD shows a consistency with no-CME scenario in the current statistics.
The method of the ongoing isobar blind analysis, and the latest sensitivity check with the event-by-event AVFD model on the different observables between Ru+Ru and Zr+Zr are also briefly discussed.
}
\maketitle
\section{Introduction}
\label{intro}
Symmetry and it's breaking are parts of the most fundamental laws in the universe. In relativistic heavy-ion collisions, accompanied with the produced hot and dense medium, one conjecture is there could be a symmetry breaking leading to the difference in the number of right-handed and left-handed quarks \cite{Kharzeev:1999,Kharzeev:2008}. This imbalance leads to a separation of electric charge in the direction of the strong magnetic field (B $\sim10^{14}$ Tesla), produced by the protons in the colliding heavy ions \cite{Mclerran:2013}. This phenomenon is known as the Chiral Magnetic Effect (CME).
 STAR (Solenoidal Tracker at RHIC) experiment is  currently the only operational experiment at the Relativistic Heavy Ion Collider (RHIC), has been contributing to the CME search over many years. 
RHIC has collided various ion species, at many different energies over the past 20 years. 
In heavy-ion collisions, CME is expected to lead to charge separation across the reaction plane determined by the impact parameter and the collision direction. The reaction plane is correlated to the direction of the magnetic field \cite{Sergei:2004}. To measure the charge separation quantitatively, one uses the well known observable known as the gamma correlator ($\gamma$).\\
\indent To reduce the charge-independent correlation backgrounds, like from global momentum conservation, we usually use the $\Delta\gamma$, which measures the difference between the correlations from the opposite charge and the same charge. There are still many background sources that have also been identified to contribute to this observable, such as the local charge conservation and the resonance decays. We can write the $\Delta\gamma$ into 2 parts, one from the  background ($\Delta\gamma_{bkg}$), another from the CME signal($\Delta\gamma_{sig}$).
In principle, the $\Delta\gamma$ can be measured with respect to any plane \cite{Haojie:2018,Sergei:2018}: e.g. using the participate plane, which is estimated by the produced particles ($\Psi_{PP}$); or the spectator plane, which is estimated by the spectators neutrons ($\Psi_{SP}$).
\section{At top RHIC energies}
\label{sec-res-top}
At the top RHIC energies, we use the Au+Au collision at 200 GeV (form year 2011, 2014, 2016) and U+U collision at 193 GeV to search for the CME. As we mentioned before, if we use the particles measured with Time Projection Chamber (TPC) to reconstruct the event plane ($\Psi_{EP}$) as a proxy of the $\Psi_{PP}$, and use the Zero Degree Calorimeters (ZDC) to reconstruct the spectator plane as a proxy of the $\Psi_{SP}$. The $\Delta\gamma$ with respect to  $\Psi_{PP}^{TPC}$ and $\Psi_{SP}^{ZDC}$ contain different fractions of the CME signal and background. If we assume the the background is propositional to the elliptic flow($v_{2}$), we will have the following relations:
\begin{linenomath*}
\begin{equation}
    \Delta\gamma(\Psi_{TPC})=\Delta\gamma^{bkg}(\Psi_{TPC})+\Delta\gamma^{sig}(\Psi_{TPC}) \\
    \label{eq:deltagamma_fcme_tpc}
\end{equation}
\begin{equation}
    \Delta\gamma(\Psi_{ZDC})=\Delta\gamma^{bkg}(\Psi_{ZDC})+\Delta\gamma^{sig}(\Psi_{ZDC}) \\
    \label{eq:deltagamma_fcme_zdc}
\end{equation}
\begin{equation}
    {\Delta\gamma^{bkg}(\Psi_{TPC})}/{\Delta\gamma^{bkg}(\Psi_{ZDC})}={v_{2}(\Psi_{TPC})}/{v_{2}(\Psi_{ZDC})} \\
    \label{eq:deltagamma_fcme_bkg}
\end{equation}
\begin{equation}
    {\Delta\gamma^{sig}(\Psi_{TPC})}/{\Delta\gamma^{sig}(\Psi_{ZDC})}={v_{2}(\Psi_{ZDC})}/{v_{2}(\Psi_{TPC})}
    \label{eq:deltagamma_fcme_sig}
\end{equation}
\end{linenomath*}
With four known variables: $\Delta\gamma(\Psi_{TPC})$, $\Delta\gamma(\Psi_{ZDC})$, $v_{2}(\Psi_{TPC})$, $v_{2}(\Psi_{TPC})$, and four unkown ones  $\Delta\gamma^{bkg}(\Psi_{TPC})$, $\Delta\gamma^{bkg}(\Psi_{ZDC})$, $\Delta\gamma^{sig}(\Psi_{TPC})$, $\Delta\gamma^{sig}(\Psi_{ZDC})$, we can solve the above equations. Then we can estimate $f_{CME}$ which represents the fraction of CME signal in the observable measured using TPC defined as: 
\begin{linenomath*}
\begin{equation}
    f_{CME}=\Delta\gamma^{sig}(\Psi_{TPC})/\Delta\gamma(\Psi_{TPC}).
    \label{eq:fcme}
\end{equation}
\end{linenomath*}
\begin{figure}[h]
\centering
\includegraphics[width=0.51\textwidth]{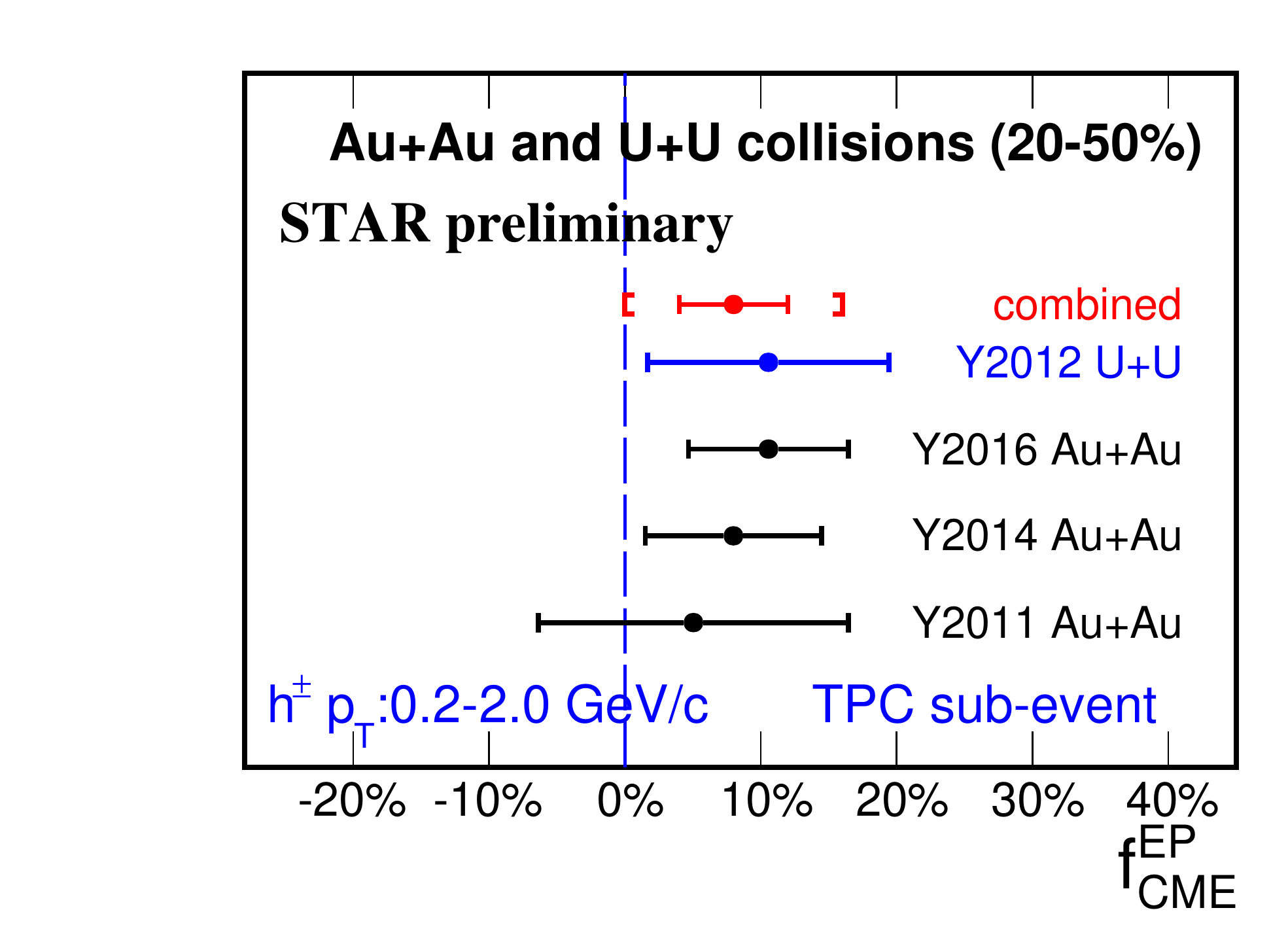}
\includegraphics[width=0.44\textwidth]{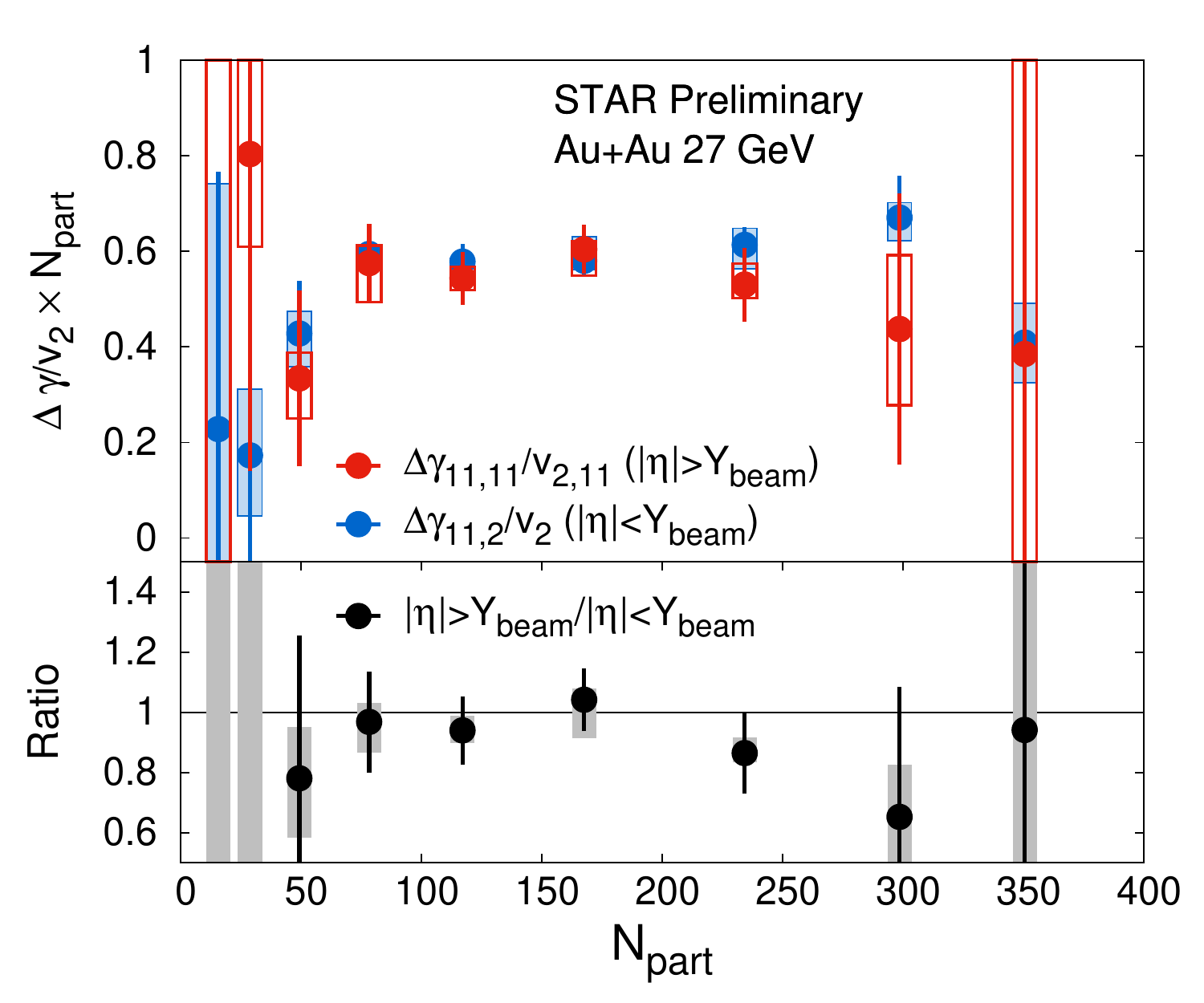}
\caption{(Left) The fraction of the possible CME signal($f_{CME}$) in $\Delta\gamma$ with respect to  TPC and ZDC event planes for Au+Au collision at 200 GeV and U+U collision at 193 GeV, combined result shows $8\pm4\pm8\%$. (Right)$\Delta\gamma$ scaled by $v_{2}$ with respect to  TPC-EPD-inner and TPC-EPD-outer planes to the no-CME scenario for Au+Au collision at 27 GeV.}
\label{fig-1}      
\end{figure}
Figure~\ref{fig-1} (left) the measurement of the $f_{CME}$ for different run years in $20\%-50\%$ centrality. The Au+Au and U+U combined result shows that the faction is $8\pm4\pm8\%$. The systematic uncertainties in this measurement are assessed by varying the track quality cuts and the $\eta$ gap. 
\section{At Lower Energy}
\label{sec-res-low}
At lower energies, the Event Plane Detector (EPD) installed in the year 2018 provides a unique capability for CME search. EPD covers a pseudorapidity range of 2.1 to 5.1 on either side of the TPC. For the $\Delta\gamma$ observable, we choose particles of interest carrying charges from the TPC, and measure the event plane from the EPD. At 27 GeV, the beam rapidity ($Y_{beam}=3.4$) falls at the center of EPD acceptance. Due to the kinematics, the inner EPD ($\eta>Y_{beam}$) detects a lot of protons, coming from spectators, stopped protons or beam fragments and they all have large directed flow which determines first-order event plane well. The outer EPD ($\eta<Y_{beam}$) detects mostly produced particles. Produced particles have large elliptic flow so we have a very well defined second order event plane. So with EPD and 27 GeV collisions we have a way to, at the same time, measure charge separation w.r.to planes of large directed flow due to spectator protons and w.r.to elliptic flow plane due to participants. \\\
\indent In the background only scenario, if the background is  propositional to $v_{2}$, the ratio of $\Delta\gamma/v_{2}$ with respect to different planes ($\Psi_{A}, \Psi_{B}, \Psi_{C}$) will be the same, in other words we expect:\\
\begin{linenomath*}
\begin{equation}
    \Delta\gamma/v_{2}(\Psi_{A})=    \Delta\gamma/v_{2}(\Psi_{B})=    \Delta\gamma/v_{2}(\Psi_{C})=...
\label{gamma:background}
\end{equation}
\end{linenomath*}
Therefore, in our case we measured the $\Delta\gamma$ and $v_{2}$ with respect to  the TPC-EPD-inner 1st order plane ($\eta>Y_{beam}$) and the $\Delta\gamma$ and $v_{2}$ with respect to  the TPC-EPD-outer 2nd order plane ($\eta<Y_{beam}$). Figure \ref{fig-1}(right) shows the $\Delta\gamma/v_{2}\times N_{part}$ with respect to  different planes. The bottom panel shows the double ratio. The results show, within the uncertainty, the ratio is consistent with unity. No significant deviation from Eq. \ref{gamma:background} is observed.
\section{Methods for blind analysis of isobar data}
\label{sec-res-isobar}
To better control the signal and the background in CME observables
isobar program is introduced. To remove unconscious human induced biases we implement the isobar blind analysis. The $\Delta\gamma$ can be written into the signal ($\Delta\gamma^{sig}$), $v_{2}$ related  background ($\Delta\gamma^{flow-bkg}$), and the non-flow driven background ($\Delta\gamma^{non-flow-bkg}$).
In an ideal scenario if we can select two different systems that produce similar non-flow effects, similar flow backgrounds, then the residual difference in charge separation can be attributed to signal -- isobar collisions were originally proposed with such expectations \cite{Sergei:2010}. In practice with the two isobar species $^{96}_{44}$Ru+$^{96}_{44}$Ru and $^{96}_{40}$Zr+$^{96}_{40}$Zr, according to model calculations one can achieve up to $4\%$ difference in  flow background \cite{weitian:2016, Schenke:2019, Haojie:prl2018, hanlin:2018}, similar non-flow  determined by the multiplicity difference, and about  B-field square difference of the order of $10-18\%$ that is relevant for a signal difference. If one can keep the systematics under control, predictions show, with two billion events collected for each species, we can achieve $5\sigma$ significance given the CME fraction in observables is at $14\%$ level \cite{weitian:2016}.\\
\indent The isobar collisions are proposed to perform an experimental test of CME with the best possible control of signal and background as compared to all previous measurements. To minimize the run-to-run variation of the detector response due to the acceptance loss, efficiency changes, or the luminosity variation, we switch the collisions of two species frequently during data taking. To minimize the unconscious bias by the analysts we: 1) use the 27 GeV data which is collected in the same year to perform a closure test of our analysis, 2) mix the Ru and Zr events to freeze the analysis methods, observables, codes, and the event selection criteria, 3) perform run-by-run quanlity assurance using blind data where species information is hidden and, 4) finally perform the full analysis after unblinding the species information. The details of this analysis method can be found in Ref. \cite{STAR:isobar}. One final cross check of the frozen code and response of different observables to CME signals has been recently performed using an event-by-event 
Anomalous-Viscous Fluid Dynamics (AVFD) simulation \cite{AVFD}. 
\section{Summary}
\label{sec-sum}
At the top RHIC energy, we measured the possible CME fraction beyond the flow background by measuring 
the $\Delta\gamma$ observable with respect to TPC and ZDC planes. At Au+Au 200 GeV and U+U 193 GeV collisions, the
combined results indicate the CME fraction is $(8\pm4\pm8)\%$ in data for $20\%-50\%$ centrality. \\
\indent At lower energy, a test of background scenario was performed using Au+Au 27 GeV data by measuring  $\Delta\gamma$ with respect to TPC and
the new installed EPD systems. The results indicate no significant deviation from background scenario with the uncertainty of our measurement. \\
\indent Meanwhile, for the ongoing analysis of the isobar collisions, a blind analysis is implemented to minimize unconscious bias. The sensitivity of different CME observables was checked using the event-by-event AVFD model. The results from the isobar blind analysis will soon be released by the STAR collaboration.

\section*{Acknowledgements}
Y. Hu is supported by the China Scholarship Council (CSC). This work was supported in part by the National Natural Science Foundation of China under contract No. 11835002.


\begin{thebibliography}{}

\bibitem{Kharzeev:2008}
Dmitri E. Kharzeev et al., Nuclear Physics A \textbf{803}, 227-253 (2008)

\bibitem{Kharzeev:1999}
Dmitri E. Kharzeev et al., Phys.Rev.\textbf{D61}, 1111901 (2000)

\bibitem{Mclerran:2013}
Larry D. McLerran. and Vladimir V. Skokov, Nucl. Phys. A \textbf{929},184-190 (2014)

\bibitem{Sergei:2004}
Sergei A. Voloshin, Phys. Rev. C \textbf{70}, 057901 (2004) 

\bibitem{Haojie:2018}
Hao-Jie Xu et al., Chinese Phys. C \textbf{42}, 084103 (2018)

\bibitem{Sergei:2018}
Sergei A. Voloshin, Phys. Rev. C \textbf{98}, 054911 (2018)

\bibitem{Sergei:2010}
Sergei A. Voloshin, Phys. Rev. Lett. \textbf{105}, 172301 (2010)

\bibitem{weitian:2016}
Wei-Tian Deng et al., Phys. Rev. C \textbf{94}, 041901 (2016)

\bibitem{Schenke:2019}
Bj\"orn Schenke et al., Phys. Rev. C \textbf{99}, 044908 (2019)

\bibitem{Haojie:prl2018}
Hao-Jie Xu et al., Phys. Rev. Lett. \textbf{121}, 022301 (2018)

\bibitem{hanlin:2018}
Hanlin Li et al., Phys. Rev. C \textbf{98}, 054907 (2018)

\bibitem{STAR:isobar}
Jaroslav Adam et al. (STAR), Nucl. Sci. Tech. \textbf{32}, 48 (2021)

\bibitem{AVFD}
Shuzhe Shi et al., Annals of Physics \textbf{394}, 50-72 (2018)

\bibitem{STAR:avfd}
Subikash Choudhury et al., arXiv:2105.06044 (2021)


\end{thebibliography}
\end{document}